%% file: main.tex
\documentclass[runningheads]{llncs}
\usepackage[T1]{fontenc}
\usepackage{booktabs}
\usepackage{graphicx}
\usepackage{multirow}
\usepackage{arydshln}
\usepackage{amsfonts}
\usepackage{amsmath}
\usepackage{xspace}
\usepackage{xcolor}
\usepackage[acronym]{glossaries}
\usepackage{adjustbox}
\usepackage[show]{chato-notes}
\usepackage{hyperref}

\DeclareFontFamily{U}{matha}{\hyphenchar\font45}
\DeclareFontShape{U}{matha}{m}{n}{
      <5> <6> <7> <8> <9> <10> gen * matha
      <10.95> matha10 <12> <14.4> <17.28> <20.74> <24.88> matha12
      }{}
\DeclareSymbolFont{matha}{U}{matha}{m}{n}

\DeclareMathSymbol{\Lt}{3}{matha}{"CE}
\DeclareMathSymbol{\Gt}{3}{matha}{"CF}

\input{src/10-macros}

\begin{document}
\title{Forward Index Compression\\for Learned Sparse Retrieval}
\titlerunning{Forward Index Compression for Learned Sparse Retrieval}


%
%
%
\author{Sebastian Bruch\inst{1}\orcidID{0000-0002-2469-8242} , Martino Fontana\inst{2} \and
Franco Maria Nardini\inst{3}\orcidID{0000-0003-3183-334X} \and
Cosimo Rulli\inst{3}\orcidID{0000-0003-0194-361X} \and
Rossano Venturini\inst{4}\orcidID{0000-0002-9830-3936}}
\authorrunning{S. Bruch et al.}
%
\institute{
Northeastern University, Boston, USA, \email{s.bruch@northeastern.edu} 
\and University of Pisa, Italy, \email{martino.fontana@studenti.unipi.it}
\and ISTI-CNR, Pisa, Italy, \email{\{name.surname\}@isti.cnr.it} 
\and University of Pisa, Italy, \email{rossano.venturini@unipi.it} 
}

\maketitle

\input{src/00-abstract}
\input{src/01-intro}
\input{src/02-related}
\input{src/03-methodology}
\input{src/04-experiments}

\input{src/05-conclusion}

\smallskip
\noindent \textbf{Acknowledgements}
This work was partially supported by the Horizon Europe RIA ``Extreme Food Risk Analytics'' (EFRA), grant agreement No. 101093026, by the PNRR - M4C2 - Investimento 1.3, Partenariato Esteso PE00000013 - ``FAIR - Future Artificial Intelligence Research'' - Spoke 1 ``Human-centered AI'' funded by the European Commission under the NextGeneration EU program, and by the MUR-PRIN 2022 ``Algorithmic Problems and Machine Learning'', grant agreement n. 20229BCXNW.

\begin{credits}
\subsubsection{\discintname}
The authors have no competing interests to declare that are relevant to the content of this article.
\end{credits}

\bibliographystyle{splncs04}
\bibliography{biblio}

\end{document}

%% file: src/10-macros.tex
\newcommand{\msmarco}{\textsc{MsMarco}\xspace}

\newcommand{\seismic}{\textsc{Seismic}\xspace}

\newcommand{\hnsw}{\textsc{Hnsw}\xspace}

\newcommand{\splade}{\textsc{Splade}\xspace}

\newcommand{\cut}{\textsf{cut}}
\newcommand{\heapfactor}{\textsf{heap\_factor}\xspace}

\newcommand{\oursshort}[1]{\sc{Pfi}}
\newcommand{\rgbshort}{RGB\xspace}
\newcommand{\basesvb}{StreamVByte\xspace}
\newcommand{\oursvb}{DotVByte\xspace}
\newcommand{\infless}{\textsc{LiLsr}\xspace}
\newcommand{\plain}{Uncompressed\xspace}
\newcommand{\zetam}{Zeta\xspace}

\newacronym[plural=\textsc{Llm}s,firstplural=Large Language Models (\textsc{Llm}s)]{llm}{\textsc{Llm}}{Large Language Model}

\newacronym{anns}{\textsc{Anns}}{Approximate Nearest Neighbor Search}

\DeclareMathOperator*{\nnz}{nnz}

%% file: src/00-abstract.tex

\begin{abstract}
Text retrieval using learned sparse representations of queries and documents has, over the years, evolved into a highly effective approach to search. It is thanks to recent advances in approximate nearest neighbor search---with the emergence of highly efficient algorithms such as the inverted index-based \seismic and the graph-based \hnsw ---that retrieval with sparse representations became viable in practice. In this work, we scrutinize the efficiency of sparse retrieval algorithms and focus particularly on the size of a data structure that is common to all algorithmic flavors and that constitutes a substantial fraction of the overall index size: the forward index. In particular, we seek compression techniques to reduce the storage footprint of the forward index without compromising search quality or inner product computation latency. In our examination with various integer compression techniques, we report that \basesvb achieves the best trade-off between memory footprint, retrieval accuracy, and latency. We then improve \basesvb by introducing \oursvb, a new algorithm tailored to inner product computation. Experiments on \msmarco show that our improvements lead to significant space savings while maintaining retrieval efficiency.

\end{abstract}

%% file: src/01-intro.tex

\section{Introduction}
\label{sec:intro}
In recent years, \glspl{llm} have been fruitfully employed to encode documents and queries into high-dimensional spaces while approximately preserving their pairwise semantic similarity. One attractive family of encoders or \emph{embedding functions} produces \emph{sparse} representations by mapping text into a space with tens of thousands of dimensions, but where most of the coordinates of the representation are zero. These techniques have gained much attention due two attributes: 1) \emph{interpretability}, as the representation produced is grounded in the vocabulary of the collection; and, 2) \emph{effectiveness}, as these techniques can heavily adopt query and document expansion so to actively counter the vocabulary mismatch problem at retrieval time.

Despite these properties, the use of sparse embeddings in text retrieval faced a practical challenge: efficiency of executing queries. Recently, however, several efforts emerged in the literature that enable fast and effective (approximate) retrieval for such embeddings, rendering the paradigm viable in practice.

Among what the literature has to offer, the highly efficient algorithms are often adaptations of existing \gls{anns} algorithms~\cite{bruch2024foundations} to sparse vectors (e.g., \hnsw~\cite{hnsw2020,ip-nsw18,kannolo}), or \gls{anns} algorithms designed specifically for learned sparse vectors (e.g., \seismic~\cite{bruch2024seismic,Bruch2024SeismicWave,bruch2025efficientsketchingnearestneighbor,ecir-seismic}, \textsc{Bmp}~\cite{mallia2024faster}, \textsc{Dsbp}~\cite{carlson2025dynamic}).
Regardless of how they organize their index or perform \gls{anns}, all such algorithms rely on a key data structure: the forward index.

The forward index is simply a mapping from document identifiers to sparse vectors. Denoting by $\mathcal{X}$ a forward index, its purpose can be succinctly stated as enabling the computation of the inner product $\langle q, x \rangle$ for a query $q$ and document $x \in \mathcal{X}$. That is why this structure is so critical to all sparse \gls{anns} algorithms: Algorithms use it to ``error-correct'' the retrieved set by re-ordering an initial set of candidates by exact inner product obtained from the forward index.

Storing the forward index is conceptually straightforward. Noting that a sparse vector $x \in \mathbb{R}^d$ has $\nnz(x) \ll d$, where $\nnz(x) = |\{x_i : x_i \neq 0\}|$ is the number of nonzero coordinates, it is more reasonable to store only the nonzero coordinates of $x$ in the forward index. One way to do that is by storing three arrays: \emph{components}, \emph{values}, and \emph{offsets}. The \emph{components} array records the positions of nonzero coordinates of all vectors in $\mathcal{X}$ sequentially, while \emph{values} similarly registers their nonzero values. The \emph{offsets} array tracks, for each vector $x^{(i)}$, where its components and values are stored within the corresponding arrays. This design allows one to store components and values in contiguous chunks of memory, thereby optimizing data access patterns.

It should be apparent that the size of the forward index is dominated by components and values: The number of bits required per component is $\lceil \log_2 d \rceil$, which is typically rounded to the nearest primitive data type such as the commonly-used $16$-bit or $32$-bit integers. Values are typically stored as $16$-bit floating point numbers, which turn out to be as effective as $32$-bit ones.

Given its role in sparse \gls{anns} algorithms and its substantial memory footprint, our work studies effective solutions for compressing the forward index. For concreteness, we conduct this study in the context of the state-of-the-art sparse \gls{anns} algorithm, \seismic~\cite{bruch2024seismic,Bruch2024SeismicWave,bruch2025efficientsketchingnearestneighbor,ecir-seismic}. Our specific contributions are as follows:
\begin{itemize}
\item We empirically evaluate a range of integer compression techniques applied to the forward index in \seismic, and analyze how each algorithm trades off size for efficiency. Our experiments reveal that StreamVByte~\cite{lemire2018stream} offers a compelling balance between compactness and decoding speed;
\item We introduce \oursvb, a new compression technique based on \basesvb. \oursvb is optimized for high-performance inner product computation---a key operation at the core of \seismic and sparse retrieval methods in general. Experimental results on \msmarco demonstrate that our proposed compression algorithm achieves a substantial reduction in memory usage while maintaining equivalent retrieval effectiveness; and, finally,
\item We provide a highly optimized implementation of \oursvb and \basesvb in Rust. Our code is made public within the \href{https://github.com/TusKANNy/seismic}{\seismic} library.
\end{itemize}


%% file: src/02-related.tex


%% file: src/03-methodology.tex

\section{Compressing the Forward Index}
\label{sec:meth}

We focus exclusively on compressing the components array where each component is a $16$-bit integer. Per the literature on integer sequence compression~\cite{CSUR21}, we represent a document by gaps between consecutive components. We begin by examining standard methods from the literature, then present our contribution.

Before presenting findings, we note that any permutation $\pi$ may be applied to the components provided the same permutation is applied to query vectors. That includes a $\pi$ that minimizes the gaps in a document's component sequence---a problem analogous to one encountered in inverted indexes where the objective is to reorder document identifiers to produce more compressible inverted lists.

A standard approach to finding $\pi$ is the Recursive Graph Bisection (\rgbshort) algorithm~\cite{dhulipala2016compressing}. It models the problem as a bipartite graph with ``query'' and ``data'' vertices, and permutes the data vertices to minimize the logarithmic gaps between consecutive neighbors in adjacency lists. The solution is obtained through recursive graph bisection, an iterative process that partitions the vertex set into two equal-sized subsets that optimize an estimate of the compressed size. In our formulation, components are data vertices and documents are query vertices. We use an open-source implementation of \rgbshort by~\cite{mackenzie2021faster}.\footnote{Obtained from \url{https://github.com/jmmackenzie/enhanced-graph-bisection}.}

\vspace{-0.2cm}
\subsection{Standard methods}

Table~\ref{tab:base_cmp} shows the performance of several standard compression techniques (with and without \rgbshort) applied to our problem on the \splade embeddings of \msmarco. It presents the average bits-per-component and the average time (``Scan Time'') required to compute the inner product between every document in \msmarco and $100$ randomly-sampled queries from the \texttt{dev-small} set. We use the random-access implementation provided by the \texttt{compressed-intvec} Rust package\footnote{\url{https://github.com/lukefleed/compressed-intvec}.} for all the methods except for \basesvb, which we implement ourselves.

\vspace{-0.5cm}
\subsubsection{Algorithms.} Before we discuss the results, let us briefly summarize two of the algorithms most relevant to this work. \textbf{VByte}~\cite{thiel1972program} encodes an integer $x$ using $L+1$ bytes, $b_0$ to $b_L$, where the most significant bit of each $b_i$ serves as a control bit indicating whether the integer terminates in $b_i$ ($0$) or continues in $b_{i+1}$ ($1$). Decompressing $x$ follows the formula $\sum_{i=0}^{L} (b_i \bmod 2^7) \cdot 2^{7i}$, where the modulo operation discards the control bits and the multiplication rescales the contribution of each byte.

\textbf{\basesvb}~\cite{lemire2018stream} extends VByte by taking a page from \textsc{varint}-GB~\cite{DBLP:conf/wsdm/Dean09} and \textsc{varint}-G8IU~\cite{DBLP:conf/cikm/StepanovGREO11}. It accelerates decompression through two key optimizations. First, when compressing an integer value $x$, it uses a $2$-bit control to record the number of bytes $x$ is compressed into. In this way, the control information for four values can be stored in a single byte, separately from the data.

This organization of the control information gives way to the second optimization: Decompression of all four values using the \texttt{\_mm\_shuffle\_epi8} SIMD instruction. To that end, it reads $16$ raw bytes into a SIMD register, and the ``permutation'' stored in the control byte shuffles them so that the register can be interpreted as an array of four $32$-bit integers. Not all $16$ bytes from the initial read may remain in the register: the raw data must be ``scrolled'' by the number of bytes that have not been discarded, which must be calculated separately.


\vspace{-0.5cm}
\subsubsection{Results.} Applying \rgbshort effectively reduces bits-per-component, with increased compression of up to $27\%$ with Elias Gamma. It also yields improved decompression performance because smaller values are faster to decompress---a phenomenon that is particularly pronounced for VByte. \basesvb has the fastest decompression speed, but its compression ratio is relatively suboptimal.

VByte's memory efficiency is rather surprising. Its application to inverted indexes, while enjoying fast decompression, achieves a compression ratio that is multiple times worse than other algorithms such as Zeta (cf. Table 11 in~\cite{CSUR21}).
\vspace{-0.3cm}



\begin{table}[t]
    \centering
    \caption{Effect of \rgbshort on four compression techniques in terms of average number of bits-per-component and time to compute inner products between a query and every document in \msmarco, averaged over $100$ randomly-sampled queries.\label{tab:base_cmp}}
    \begin{tabular}{lrrrr}
        \toprule
        \multirow{2}{*}{Method} & \multicolumn{2}{c}{Bits per Component} & \multicolumn{2}{c}{Scan Time (sec.)} \\
        \cmidrule(lr){2-3} \cmidrule(lr){4-5}
        & No \rgbshort & \rgbshort & No \rgbshort & \rgbshort \\
        \midrule
        \plain & 16.0 & - & 0.3 & - \\
        \midrule
        VByte~\cite{thiel1972program} & 11.8 & 9.6 & 9.9 & 7.1 \\
        Elias Gamma~\cite{elias75} & 12.1 & 8.6 & 10.1 & 9.6 \\
        Elias Delta~\cite{elias75} & 10.9 & 8.1 & 7.9 & 7.6 \\
        Zeta~\cite{boldi2005codes} & 9.8 & 7.5 & 8.4 & 7.1 \\
        \basesvb~\cite{lemire2018stream} &12.5 & 11.2 & 1.3 & 1.2 \\
        \bottomrule
    \end{tabular}
    \vspace{-4mm}
\end{table}

\subsection{\oursvb: our contribution}



We now describe how we specialize \basesvb for the components array compression problem. First, observe that components are $16$-bit integers, implying that a single control bit is adequate. As a result, \texttt{\_mm\_shuffle\_epi8} can decompress $8$ values simultaneously into a $128$-bit register. In addition, the number of raw bytes to scroll is already encoded in the control byte itself through \texttt{popcnt}.

The second difference lies in the implementation. \basesvb is general-purpose and stores the decoded values into a buffer before reading them individually. In contrast, we design \oursvb for maximum query performance by executing a sequence of operations directly in SIMD registers to compute inner products: a) decompress the components; b) gather the query values in the corresponding positions; 
c) read document values, convert them to \texttt{f32}, multiply them with the gathered query values, and add the packed product to the result.

Finally, in \basesvb, a control byte can be shared between two consecutive documents. As a result, we are forced to decompress some values from the previous document, introducing unwanted overhead. We, however, enforce per-document alignment: because the number of components encoded is a multiple of $8$, all remaining components are stored uncompressed and processed normally.

%% file: src/04-experiments.tex

\section{Experimental Evaluation}
\label{sec:exp}

We now evaluate \oursvb and baselines on \msmarco encoded using two state-of-the-art sparse embedding models: 1) \splade~\cite{cocondenser}, widely used in retrieval efficiency benchmarks, and 2) Learned Inference-Free (\infless)~\cite{nardini2025effective}.

\infless belongs to a family~\cite{DBLP:journals/corr/abs-2411-04403,10.1145/3726302.3730192} that forgo query embedding and assign precomputed scores to query terms. The absence of query expansion, however, necessitates greater document expansion~\cite{nardini2025effective,DBLP:journals/corr/abs-2411-04403,10.1145/3726302.3730192}: \splade embeddings of \msmarco have $119$ ($43$) nonzero terms per document (query) while \infless embeddings have $387$ ($6$). This $3.2\times$ increase in the number of nonzero terms presents a challenge to compression and inner product computation efficiency.

\smallskip
\noindent \textbf{Application to \seismic}.
\seismic is an efficient \gls{anns} algorithm designed for sparse embeddings. It works by organizing inverted lists into blocks, each with its own ``summary'' vector. During query processing, blocks are evaluated using the summary vector and, if the block is flagged, the exact inner product between the query and every document in that block is computed using the forward index.

We test compression algorithms using the publicly available implementation of \seismic.\footnote{\url{https://github.com/TusKANNy/seismic}} For each dataset, we select the best index with the constraint that the overhead caused by the inverted index and summaries does not exceed $50\%$ the size of the collection---reported as memory budget of $1.5\times$ in~\cite{Bruch2024SeismicWave,ecir-seismic}. On the resulting index, we tune query hyperparameters: \heapfactor $\in \{ 0.7, 0.8, 0.9, 1.0\}$ and \cut $\ \in \{2, 4, 6, 7, 8, 10, 12 \}$.

\smallskip
\noindent \textbf{Hardware}. Our code is compiled using Rust 1.94.0 (nightly). Experiments were conducted on a server equipped with an Intel(R) Core Ultra 7 CPU with 124 Gbyte of RAM with 20 cores. Search is performed using a single thread.

\smallskip
\noindent \textbf{Results}. 
Our experiments indicate that the scan time of \oursvb is $0.36$ seconds (merely $20\%$ slower than \plain) and that it yields $10.8$ bits per component under the same conditions of Table~\ref{tab:base_cmp} (with \rgbshort). In comparison to \basesvb, it is more than $3 \times$ faster with an even smaller memory footprint.

Table \ref{tab:cmp_combined} summarizes our results by average query latency ($\mu$sec.) at four different accuracy levels (i.e., the percentage of true nearest neighbors recalled), three different compression techniques for components, and using \texttt{float16} and \texttt{fixedU8}\footnote{With the \texttt{fixed} Rust package: \url{https://docs.rs/fixed/latest/fixed/}.} representation for values. While our contribution focuses on component compression, we show that it is possible to halve the storage space required for the values with minimal performance degradation. Additionally, we report the space needed (in GB) to store the inverted index and forward index.


Among compressed indexes, \zetam yields the highest query latency but the largest reduction in space usage. \basesvb trades off space for speed, with an average query latency which is about three times greater than \plain.

\oursvb outperforms others in the latency-size trade-off space. In fact, it achieves a space reduction of up to $22\%$ over \plain ($5\%$ to $10\%$ overhead with respect to \zetam) with similar query latency, both for \splade and \infless. 


\begin{table}[t]
    \centering
    \caption{Mean latency ($\mu$sec) and index size (GB). Values are \texttt{float16} or \texttt{fixedU8}---with the latter, no solution reaches 99$\%$ accuracy given the memory budget.
    \label{tab:cmp_combined}}
    \adjustbox{max width=\textwidth}{
    \begin{tabular}{lrrrrrrrrrr}
        \toprule
        \splade  &\multicolumn{5}{c}{\texttt{float16}} & \multicolumn{5}{c}{\texttt{fixedU8}} \\ 
        \cmidrule(lr){2-6} \cmidrule(lr){7-11}
        Accuracy (\%)&  90 & 95 & 97 & 99 & GB & 90 & 95 & 97 & \,99\, & GB \\ 
        \midrule
        \plain & 157 & 289 & 387 & 1,081 & 5.8 &  185 & 290 & 485 &  - & 4.8 \\

        \midrule
        \zetam &  848 & 1,616 & 2,343 & 6,243 & 4.8 & 825 & 1,702 &2,881 & - & 3.8  \\ 
        \basesvb & 425 & 818 & 1168 & 3146 & 5.2 & 402 &  768 & 1288 & - & 4.2\\ 
        \oursvb & 165 & 309&  412 & 1082& 5.1  &156 & 305 & 482 & - & 4.2 \\
        \midrule
        \midrule
        \infless & \multicolumn{5}{c}{\texttt{float16}} & \multicolumn{5}{c}{\texttt{fixedU8}} \\ 
        \cmidrule(lr){2-6} \cmidrule(lr){7-11}
        Accuracy (\%) & 90 & 95 & 97 & 99 & GB & 90 & 95 & 97 & 99 & GB\\ 
        \midrule
        \plain & 518 & 773 & 773& 1,658 & 17.5 & 670 & 734 & 734 & - & 14.3  \\ 
        
        \midrule
        \zetam  &  8,220 & 12,093 & 12,093 & 25,401 & 13.2 & 11,363 & 11,889 & 13,196 & - & 10.1\\
        \basesvb & 2,019 & 3,123 & 3,123 & 6,493 & 15.4 & 2,386 &2,559 & 2,943 & - & 12.2 \\ 
        \oursvb & 529 & 788 & 788 & 1,643 &14.9 & 662 & 728 & 822 & - &11.7 \\
        \bottomrule
    \end{tabular}}
    \vspace{-0.4cm}
\end{table}

%% file: src/05-conclusion.tex
\section{Conclusions and Future Work}
\label{sec:conclusion}
\vspace{-0.2cm}
We examined the efficiency of sparse \gls{anns} algorithms with a particular focus on the forward index---a substantial contributor to index size. By exploring a suite of integer compression techniques, we identified \basesvb as providing the most favorable balance between memory usage, retrieval accuracy, and latency.

We then introduced an enhanced variant of \basesvb, dubbed \oursvb, that is specifically optimized for inner product. Experiments on \msmarco confirm that it achieves significant space savings while preserving search quality and efficiency. These results highlight the potential of targeted compression strategies to improve the practicality of learned sparse retrieval systems.

For future work, we plan to incorporate sub-byte capability into \oursvb for small, frequent dgaps, to further improve the compression ratio.